\documentclass[12pt]{article}
\usepackage{epsfig}

\begin{document}

\begin{center}
{\large {\bf Coupled Ising models with disorder}}\\
\vspace{1cm}
{\bf P. Simon$^a$ and F. Ricci-Tersenghi$^b$}\\
\vspace{0.5cm}
$^a${\em International School for Advanced Studies, Via Beirut 2-4,
34013 Trieste, Italy} \\
$^b${\em Abdus Salam International Center for Theoretical Physics,
Condensed Matter Group}\\
{\em Strada Costiera 11, P.O.Box 586, 34100 Trieste, Italy}\\
\end{center}
\vspace{0.2cm}
\begin{abstract}
\noindent                       
In this paper we study the phase diagram of two Ising planes coupled
by a standard spin-spin interaction with bond randomness in each
plane.  The whole phase diagram is analyzed by help of Monte Carlo
simulations and field theory arguments.
\end{abstract}
\vspace{0.5cm}

\section{Introduction}

In order to understand the role of impurities and inhomogeneities in
real physical systems, many statistical models with quenched
randomness have been proposed. The effect of randomness on continuous
phase transitions has thus been of great interest for many years.  The
main question is to determine whether the randomness leaves the
critical properties of the pure system unchanged. One prediction
concerning models with random bonds comes from the Harris
criterion~\cite{harris}, which states that bond randomness changes the
values of the critical exponents only if the specific heat exponent
$\alpha$ is positive.  The effects of bond randomness have been
studied intensively in the $2D$ Ising model and more recently in the
$q<4$ states Potts model. In the former case, the $log$ singularity of
the specific heat is transformed in a $log(log)$
singularity~\cite{dots83}, whereas in the latter case a new critical
point with new critical exponents is reached~\cite{dots95}.
Furthermore, it has been recently realized that a weak bond randomness
can also have strong effects on $2D$ spin systems possessing a first
order transition. Following earlier work of Imry and
Wortis~\cite{imry}, Hui and Berker~\cite{hui} have shown, using a
phenomenological renormalization group, that bond randomness (even
infinitesimal) induces a second order transition (i.e.\ vanishing
latent heat) in a system that would have undergone a first order
one. This result has been proved rigorously by Aizenman and
Wehr~\cite{aizen}. At this level, a question naturally arises
concerning the universality class of this second order transition. In
order to test these predictions, the $q$ states Potts model ($q>4$ is
necessary to get a first order transition) with bond randomness has
been recently extensively analyzed numerically~\cite{potts}, the
conclusion being that the transition is of second order type due to
disorder and the critical behaviour depends on $q$.  Analytical
results on the $q$ states Potts model are, as usual, much more
difficult to obtain. However, this is not the case for those systems
presenting what Cardy called a ``weak fluctuation-driven first order
transition''~\cite{cardy95}.  The typical example Cardy provided is a
system of $N$ critical Ising models coupled by their energy
density. Without disorder, this perturbation drives the system in a
strong coupling regime where perturbative renormalization group (RG)
analysis fails. Yet, for this model, it can be shown
non-perturbatively that this runaway flow is indeed associated to a
massive regime (with a finite correlation length depending on the
coupling). This is a way to define this ``weak fluctuation-driven
first order transition'' phenomenon. The addition of weak bond
randomness, even infinitesimal, makes the system flow to $N$ decoupled
critical Ising models~\cite{cardy95}.  This result has been extended
to the case of $N$ coupled Potts models~\cite{pujol,simon}.  The
result was a non-Ising like second order transition, the universality
class depending on the sign of the coupling between the models. This
class of examples is interesting because some infinitesimal bond
randomness is able to change the ``weak fluctuation-driven first order
transition'' into a continuous one. Concretely speaking, the RG runaway flow is completely turned over by disorder preventing a strong coupling regime. Nevertheless a lot of questions
still remain open. What happens if we take a different coupling than
the energy density one?  Does infinitesimal disorder will also change
the order of the transition?

To give some insights to these questions, we would like to study the
following system: two Ising planes coupled by a spin-spin interaction
where disorder (dilution or bond-randomness) is added in each
plane. When the two Ising planes are critical, the spin-spin
perturbation which is much stronger that the energy-energy
perturbation drives the system in a strong coupling regime. It has
been shown using integrable Toda field theories that a finite
correlation length is generated~\cite{leclair}, and the full spectrum
of the theory can be computed too. In this sense, the situation is
analogous to the model treated by Cardy~\cite{cardy95}, except that
now the coupling is different and much stronger. Therefore, this model
can bring some clues to the aforementioned questions.  Moreover, it is
worth noticing that $N$ Ising planes (with $N \gg 1$) coupled by the
spin-spin interaction is a way to reach a $3D$ Ising model.  $N=2$ may
be regarded as a kind of first step in this direction.
Therefore, we expect Aizenman-Wehr theorem not to apply here.

The plan of the paper is the following. In section 2, we present
analytical field theory arguments showing that a finite amount of
disorder is necessary to change the ``weakly fluctuation-driven order
first order transition'' into a continuous one. In section 3, we
confirm it by Monte Carlo numerical simulations. In section 4, we will
summarize the results and discuss the whole phase diagram of the
model.

\section{Analytical results}

\subsection{Pure case}

In this section, we first summarize results available in the pure
case. The Hamiltonian then reads
\begin{equation}
\label{ham}
H = - J \sum_{<i,j>} (\sigma_i^1 \sigma_j^1 + \sigma_i^2 \sigma_j^2) -
J' \sum_i \sigma_i^1 \sigma_i^2 \quad ,
\end{equation}
where $\sigma_{i}^{1,2}=\pm 1$, the first sum is over the nearest
neighbours on a 2D square lattice, the same coupling $J$ has been
chosen for the two models and a coupling $J'$ has been considered
between the two planes.  This model can be described in the continuum
limit by the following action
\begin{equation}
\label{cft}
{\cal A} = {\cal A}_{Is}^1 + {\cal A}_{Is}^2 + m \int {\rm d}^2 x
(\varepsilon_1(x)+\varepsilon_2(x)) + g \int {\rm d}^2 x~ \sigma_1(x)
\sigma_2(x) \quad ,
\end{equation}
where ${\cal A}_{Is}^i$ denotes the action of the critical Ising
model, $\epsilon_i(x)$ the Ising energy operator, $\sigma_i$ the Ising
spin operator, $m \propto (T-T_c)$ and $g \propto J'$.

When $g=0$, the action correspond to two decoupled massive Ising
models which can be described by free massive Majorana fermions or by
free Dirac fermions or by a sine-Gordon model. If the strongly
relevant coupling $g$ is switched on at the $m=0$ critical point, the
system is known to be driven into a massive regime whose mass spectrum
has been computed exactly~\cite{leclair}.  When $g\ne 0$, the bosonic
representation is more suitable (in terms of Majorana fermions
representation, this interaction is indeed non-local), and the
generalized sine-Gordon reads
\begin{equation}
\label{action}
{\cal A} = \int {\rm d}^2 x \frac12(\nabla \phi)^2 + m \cos(\beta\phi)
+ g \cos(\alpha\phi\pm\delta) \quad ,
\end{equation}
with $\beta=\sqrt{4\pi},~\alpha=\sqrt{\pi}$ and $\delta={\pi\over 2}$.

In order to obtain this expression, we write the Hamiltonian in terms
of Majorana fermions and then use standard bosonization formula (see
for example~\cite{gnt} for a review). It is worth noticing that the
addition of a four spin interaction like $K \sum\limits_{<i,j>}
\sigma_i^1\sigma_j^1\sigma_i^2\sigma_j^2$ in the original Hamiltonian
(\ref{ham}) (which therefore defines a generalized Ashkin-Teller
model) would not change the form of the action (\ref{action}) except
the substitution $(\sqrt{4\pi},\sqrt{\pi}) \longrightarrow
(\beta,\frac12\beta)$ with $\beta=\beta(K)$.

The action (\ref{cft}) is {\it a priori} hard to study since it
contains two strongly relevant perturbations and it is unfortunately
non integrable.  Notice that $g$ is much stronger that $m$ and scales
around the critical point like $g\sim |m|^{7/4}\sim |T-T_c|^{7/4}$. A
perturbed conformal field theory approach is unappropriate and beyond
feasibility because the flow is quickly driven in a strong coupling
regime.

\begin{figure}
\centerline{\epsfig{file=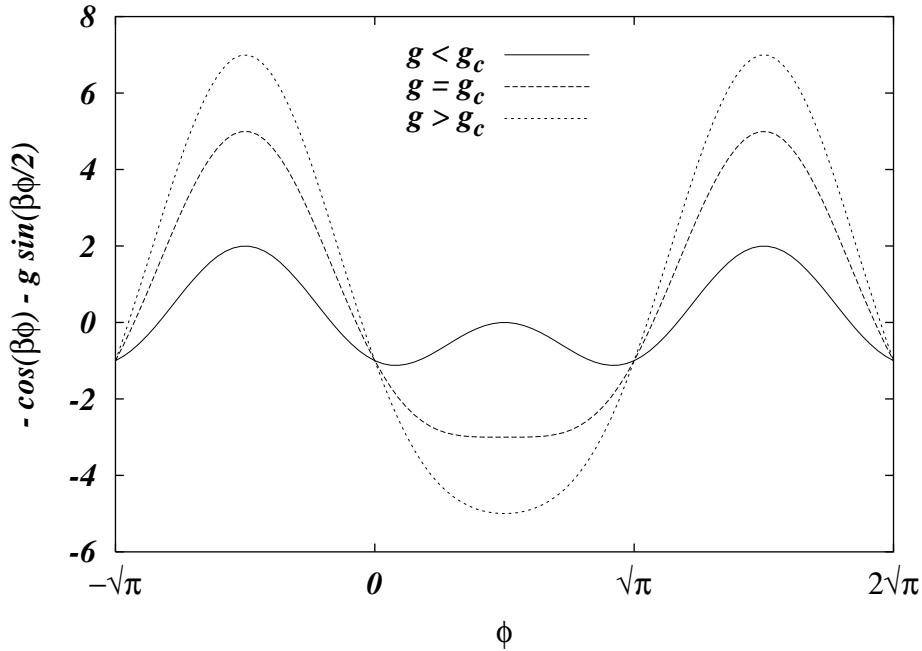,width=0.8\textwidth}}
\caption{Evolution of the potential when varying $g$ at fixed mass $m$.
Criticality is reached for a critical value $g_c$.}
\label{pot}
\end{figure}

The action (\ref{action}) which is one particular case of the double
frequency sine-Gordon model has been qualitatively studied in the
general case by Delfino and Mussardo~\cite{dm}.  It has been argued by
looking at the solitonic structure in both weak and strong coupling
regime ($m \gg g$ and $m \ll g$) that when the ratio ${\alpha\over
\beta} \equiv {n' \over n}$ is a rational number and $\delta
={\pi\over n}$, a phase transition should occur for $m>0$ ({\it i.e.}
$T>T_c$). Concerning our particular case, we can easily convince
ourselves about this transition by looking at the evolution of the
shape of the potential ${\cal V}=-m\cos(\beta\phi)-g\sin(\beta\phi/2)$
which passes from a periodic two degenerate minima to a periodic one
absolute minimum situation when increasing the ratio $g/m$ (see Figure
\ref{pot}).  From a Ginzburg-Landau point of view, it suggests that
the transition is in the universality class of $\Phi^4$, namely is of
Ising type.

Let us come back to the original Hamiltonian (\ref{ham}). In the limit
$g \to \infty$, the spins $\sigma_i^1$ and $\sigma_i^2$ are locked to
the same value and the model reduces to only one Ising model with a
critical temperature $2J_c$.  For a generic value of $g$ the same
scenario should occur for a critical temperature $J_c<T_c(g)<2J_c$
since $g$ tends to order the system in contrast to
temperature. Therefore, there should be a critical line in the $(T,g)$
plane interpolating between the fixed points $(J_c,0)$ and
$(2J_c,\infty)$.  In the $g \gg J$ limit, we can rewrite the
Hamiltonian (\ref{ham}) in the basis
$\sigma_i=\sigma_i^1;\tau_i=\sigma_i^1\sigma_i^2$ and integrate over
the $\tau$ variable using a mean field approximation (we refer to
\cite{fgn} for a more detailed analysis in a different context). It
remains {\it one} Ising model with approximate effective coupling
$2J(1-c{J^2\over g^2})\approx 2J_{\rm eff}$, with $c$ a positive
number satisfying a self-consistent equation. The Ising criticality is
reached when $J_{\rm eff}=J_c$. This analysis is valid only close to
the $(2J_c,\infty)$ Ising fixed point, namely in the strong coupling
limit.

According to Zamolodchikov c-theorem~\cite{zamolod} (without disorder
the theory is still unitary), the flow goes from the $(J_c,0)$ fixed
point with central charge $c=1$ toward the ending Ising fixed point
$(2J_c,\infty)$ with central charge $c=\frac12$. From the continuum
limit point of view, it is worth noticing that the elementary
excitations around this fixed point have nothing to do with the ones
around the $c=1$ fixed point (the correspondence is indeed non local
explaining why a standard perturbation theory around the $c=1$ fixed
point is hopeless).

\subsection{Disordered case}
Let us now add bond randomness or dilution in each Ising planes.  In
the two limits $g=0$ and $g=\infty$ the problem reduces to the well
known disordered Ising model and there are no new fixed points.  Can
we say something in the general situation?  The phase diagram is
parametrized by three variables, $T,~g$ and $\Delta$ the strength of
the disorder. In the planes $\Delta=0$ and $g=0$, the shape of the
transition separating the ferromagnetic phase from the paramagnetic
one is known respectively from results concerning the disordered Ising
model and from the pure case analysis described above. Therefore, by a
continuity arguments, we expect a critical surface joining the two
Ising critical lines.  Let us focus on the $T=T_c$ plane.  In the
continuum limit, the effect of disorder is to change $m$ into $m(x)$
in the action (\ref{action}) with $\overline{m(x)}=0$, $\overline{m(x)
m(x')}=\Delta \delta(x-x')$ for a Gaussian disorder. This action is
very hard to study with usual tools and we can only give qualitative
arguments.  At fixed value for $g$, when $m(x)>m^*(g)=g/4$, it implies
a local high temperature region surrounded by low temperature
regions. In the limit $\Delta \gg m^*(g)$ (still at fixed $g$), the
phenomenology of the critical disordered Ising model is recovered with
a mixture of low and high temperature regions. This suggests that
there should exist a finite critical value $\Delta^*(g)$ of the
disorder strength able to disorder the system. In fact, the situation
resembles the one of a disordered Ising model with
$\overline{m(x)}=m_0>0$. This is clearly a non-perturbative
phenomenon. We conjecture this transition to be of Ising type governed
by the Ising fixed point (with $c={1\over 2}$) at
$(2J_c,\infty,\Delta=0)$ with a $log(log)$ behavior for the specific
heat due to the disorder. Another non trivial fixed point at
$\Delta^*>0$ cannot {\it a priori} be excluded but is very unprobable
according to the symmetries of the model under consideration. Note
that contrary to the examples treated in~\cite{cardy95,pujol,simon}
where some infinitesimal disorder changes the weakly driven first
order transition into a continuous one, here a {\it finite} value of
the disorder is needed!

\section{Numerical analysis}
To put on firmer grounds these qualitative arguments, we have
performed Monte Carlo simulations at $T=T_c$. Let us present our
numerical results.  We have considered two $2D$ Ising models, with
bond dilution $\Delta$ in both planes, i.e.\ $P(J) = \Delta \delta(J)
+ (1-\Delta) \delta(J-1)$.  The disorder is independent in each plane.
We let the planes interact via a spin-spin coupling term, whose
strength is $g$.

\begin{figure}
\centerline{\epsfig{file=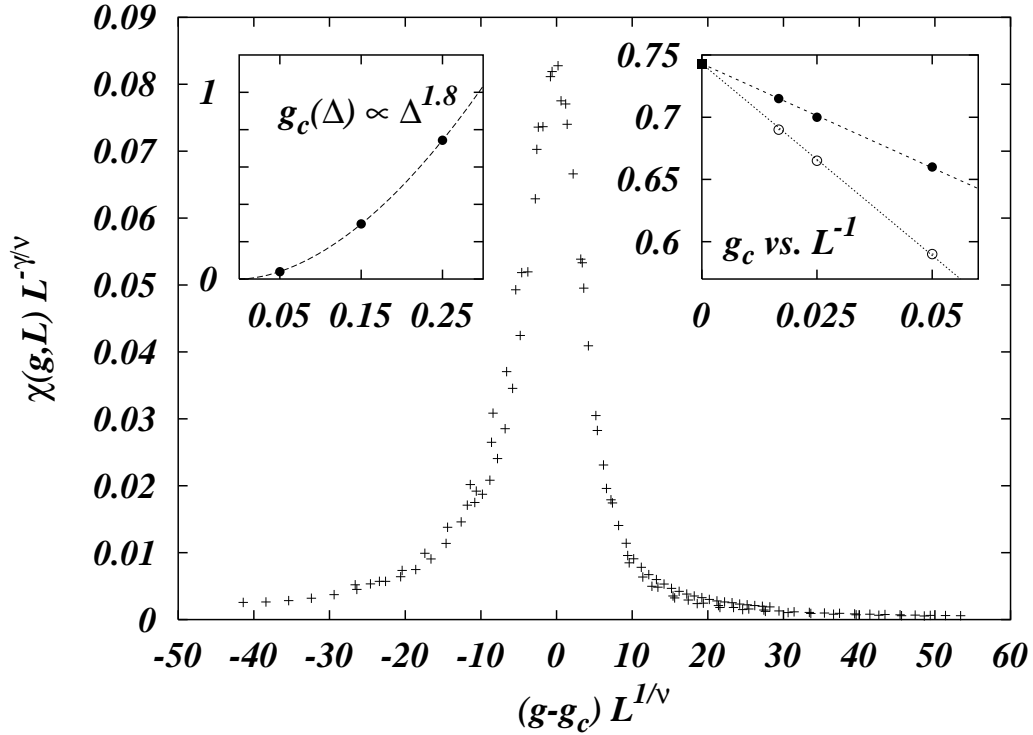,width=0.9\textwidth}}
\caption{Susceptibility data rescaled with pure $2D$ Ising model
exponents: $\nu=1$ and $\gamma=7/4$. In the left inset we report the
values for the critical coupling $g_c$ versus the disorder amount
$\Delta$, together with the best power law fit. In the right inset we
show, as a function of $1/L$, the $g$ values where the specific heat
(filled symbols) and the susceptibility (empty symbols) have a
peak. Here $\Delta=0.25$ and the Binder cumulants crossing point is
$g_c=0.743$ (filled square).}
\label{plot}
\end{figure}

We have performed numerical simulations with 3 different lattice sizes
($L=20,40,60$) and for three values of the disorder:
$\Delta=0.05,0.15,0.25$.  Lattice sizes may seem small with respect to
previous studies on the disordered Ising model~\cite{selke}, however they seem to be in the scaling
regime (see below) and so the extrapolation to large sizes is safe, at
least up to the precision we are interested in.  For any disorder
value we have found clear evidences that, increasing the value of the
coupling $g$, the system undergoes a phase transition from a
paramagnetic to a ferromagnetic phase.  The critical coupling values
can be well fitted by a power law: $g_c(\Delta) \propto \Delta^{1.8}$
(see left inset in Fig.~\ref{plot}, where the errors are smaller than
the symbol size).  Note that the value of this exponent is very close
to the theoretical one obtained by scaling arguments close to the
$c=1$ fixed point, which is $7/4$.

For every $\Delta$, the critical coupling $g_c$ has been determined as
the crossing point of the Binder cumulants. This crossing point has
almost no finite size effects.  Two more estimations of $g_c$ can be
obtained from the $g$ values where the specific heat and the
susceptibility have a peak. In fact these values converge to $g_c$ in
the large $L$ limit.  However they show much larger finite size
effects than the Binder cumulants crossing point.  Then, in order to
have a better estimation of $g_c$, we have used the data for every
lattice size and we have considered finite size corrections through
the formula $g_c(L) = g_c - A L^{-1/\nu}$, with $\nu=1$.  We have
found that our data are compatible with that formula for any disorder
value and the $g_c$ estimation coincides with the one extracted from
the Binder cumulants (see right inset in Fig.~\ref{plot}).  The effect
of increasing $\Delta$ simply reflects in a larger value for the
non-universal coefficient $A$.

In order to verify that the transition is governed by the pure $2D$
Ising model fixed point, we have analyzed our numerical data using
finite size scaling and the critical exponents of the pure $2D$ Ising
model, i.e.\ $\nu=1$ and $\gamma=7/4$.  We have found very good data
collapse for any disorder value $\Delta$.  In particular in
Fig.~\ref{plot} we show the collapse of the rescaled susceptibility
for the larger disorder value $\Delta=0.25$.  For smaller disorder
amplitudes the scaling is even better and the scaling function is more
peaked signaling that the critical region is narrower.

\section{Discussion and conclusions}

\begin{figure}
\centerline{\epsfig{file=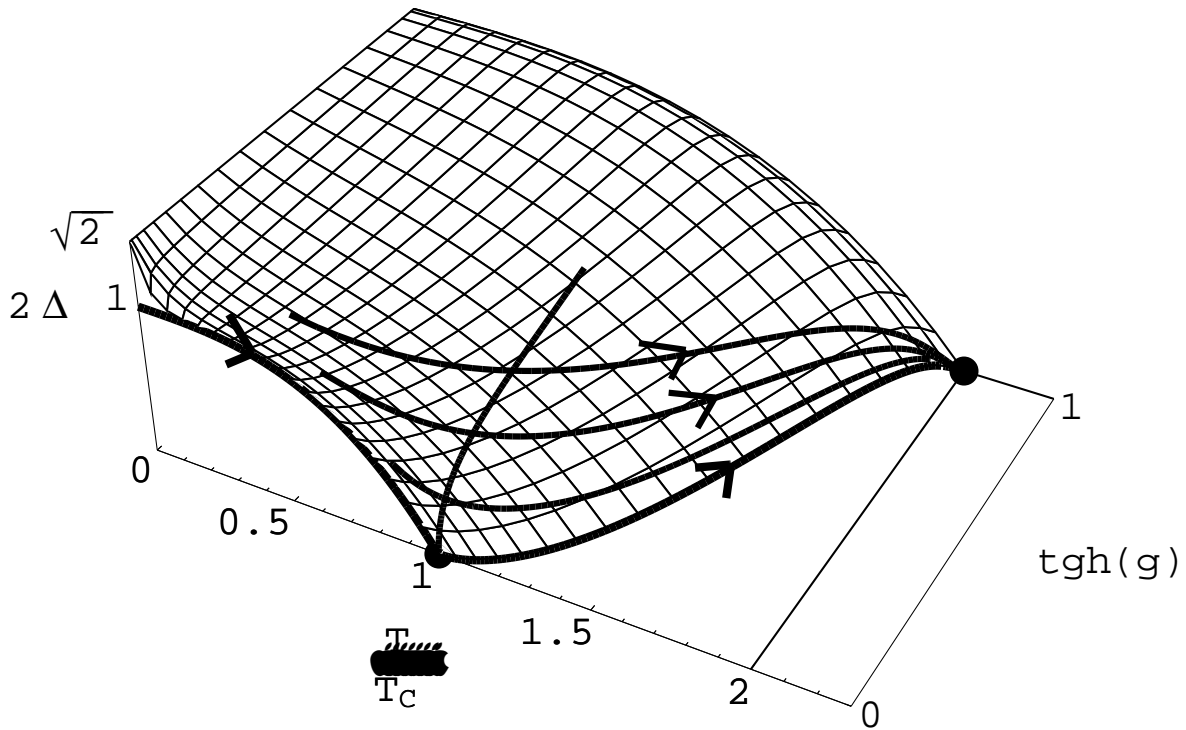,width=\textwidth}}
\caption{Phase diagram in the variables $T,g,\Delta$, where $\Delta$
is the bond dilution. The critical surface separates the ferromagnetic
region (below) from the paramagnetic one (above). The arrows indicate
the flow direction. We have numerically studied the system along the
line $T=T_c$ shown in the figure.}
\label{pd}
\end{figure}

In this paper we have presented analytical and numerical evidences
that in two coupled 2D Ising models the ``weak fluctuation-driven first order
transition'' changes to a continuous one only when a finite amount of
disorder is added.  Moreover, we have seen that even for finite and
large disorder amplitudes, the phase transition is in the  $2D$ 
disordered Ising  universality class (Ising transition modified by logarithm corrections).  The results we get can be summarized
in the schematic $3D$ phase diagram depicted in Fig.~\ref{pd}.  Our
analysis strongly suggests that almost all the critical surface shown
in Fig.~\ref{pd} is governed by the $2D$ Ising fixed point, except
the curve in the $g=0$ plane (governed by the $c=1$ fixed point) and
the one in the $T=0$ plane (governed by a percolation fixed
point\footnote{Note that the $T=0$ critical line is discontinuous at
$g=0$, as can been seen in Fig.~\ref{pd}. In fact for every $g \neq 0$
the bond percolation threshold is $1/\sqrt{2}$, while for $g=0$ is
$1/2$}).  Because of the known flow lines on this surface (the ones
with an arrow in Fig.~\ref{pd}), we believe that any transition across
the surface is governed by the $c=1/2$ fixed point located in $(2
J_c,\infty)$ (the rightmost big dot in Fig.~\ref{pd}) because the
$c=1$ fixed point is strongly repulsive in the $g$ direction (even if
it is attractive on the $g=0$ plane~\cite{dilu2d}).  For an even
clearer numerical check, a numerical estimation of the central charge
would be wellcome.

As a conclusion, we would like to mention that several directions
remain open. First of all, a theoretical framework to describe
quantitatively such non-pertubative transition is needed. Secondly it
would be nice extending this analysis for $N>2$ coupled Ising models
and finally it is worth performing a similar analysis for correlated
disorder.

\vskip 0.3cm {\bf Acknowledgements} One of us (PS) would like to
acknowledge interesting discussions with Vl. S. Dotsenko, M.-A. Lewis,
G. Mussardo and A. A. Nersesyan. This work has been partially supported by the EC
TMR Programme {\em Integrability, non--perturbative effects and
symmetry in Quantum Field Theories}, grant FMRX-CT96-0012.

\end{document}